\documentstyle[aps,prl,epsfig,multicol]{revtex} 
\def\vecgr#1{\mathchoice{\mbox{\boldmath$\mathrm\displaystyle#1$}}
{\mbox{\boldmath$\mathrm\textstyle#1$}}
{\mbox{\boldmath$\mathrm\scriptstyle#1$}}
{\mbox{\boldmath$\mathrm\scriptscriptstyle#1$}}}

\def\vec#1{\mathchoice{\mathrm{\mathbf{\displaystyle#1}}}
{\mathrm{\mathbf{\textstyle#1}}}
{\mathrm{\mathbf{\scriptstyle#1}}}
{\mathrm{\mathbf{\scriptscriptstyle#1}}}}
\newcommand{\bi}{\begin{itemize}}
\newcommand{\ei}{\end{itemize}}

\newcommand{\be}{\begin{equation}}
\newcommand{\ee}{\end{equation}}
\newcommand{\ba}{\begin{array}}
\newcommand{\ea}{\end{array}}
\newcommand{\bea}{\begin{eqnarray}}
\newcommand{\eea}{\end{eqnarray}}

\newcommand{\bc}{\begin{center}}
\newcommand{\ec}{\end{center}}
\newcommand{\bslide}{\begin{slide}}
\newcommand{\eslide}{\end{slide}}

\newsavebox{\TRS}
\sbox{\TRS}{\hspace{.5em} = \hspace{-1.8em}
                 \raisebox{1ex}{\mbox{\scriptsize TRS}} }

\newsavebox{\defgleich}
\sbox{\defgleich}{\ :=\ }

\newsavebox{\LSIM}
\sbox{\LSIM}{\raisebox{-1ex}{$\ \stackrel{\textstyle<}{\sim}\ $}}
\newcommand{\lsim}{\usebox{\LSIM}}
\newsavebox{\GSIM}
\sbox{\GSIM}{\raisebox{-1ex}{$\ \stackrel{\textstyle>}{\sim}\ $}}
\newcommand{\gsim}{\usebox{\GSIM}}

\newcommand{\rar}{\rightarrow}

\newcommand{\lk}{\left}
\newcommand{\rk}{\right}
\newcounter{saveeqn}

\newcommand{\sssty}{\scriptscriptstyle}
\newcommand{\fns}{\footnotesize}

\newcommand{\ra}{\,\rangle}
\newcommand{\la}{\langle\,}

\newcommand{\ptdnd}[1]{\frac{\partial}{\partial #1}}

\newcommand{\eh}{\mbox{$\frac{1}{2}$}}
\newcommand{\edzi}{\mbox{$\frac{1}{2i}$}}
\newcommand{\ev}{\mbox{$\frac{1}{4}$}}

\newcommand{\ih}{\mbox{$\frac{i}{2}$}}

\newcommand{\iohbar}{\frac{i}{\hbar}}

\newcommand{\intdd}[1]{\int\,d^3\!#1\,}

\newcommand{\alp}{\alpha}

\newcommand{\gam}{\gamma}

\newcommand{\del}{\delta}
\newcommand{\Del}{\Delta}

\newcommand{\ome}{\omega}
\newcommand{\Ome}{\Omega}

\newcommand{\kbar}{{\bar k}}

\newcommand{\oa}{\hat{a}}

\newcommand{\oadg}{\hat{a}^\dagger}
\newcommand{\oatdg}{\tilde{a}^\dagger}
\newcommand{\oat}{\tilde{a}}

\newcommand{\oal}{\hat{\alp}}
\newcommand{\oaldg}{\hat{\alp}^\dagger}

\newcommand{\obdg}{\hat{b}^\dagger}
\newcommand{\ob}{\hat{b}}

\newcommand{\obet}{\hat{\beta}}

\newcommand{\obetdg}{\hat{\beta}^\dagger}
\newcommand{\obettil}{{\tilde{\beta}}}
\newcommand{\obettildg}{{\tilde{\beta}}^\dagger}

\renewcommand{\oc}{\hat{c}}
\newcommand{\ocd}{\dot{\hat{c}}}
\newcommand{\ocdg}{\hat{c}^\dagger}

\newcommand{\of}{\hat{f}}
\newcommand{\ofdg}{\hat{f}^\dagger}
\newcommand{\oF}{\hat{F}}
\newcommand{\oFdg}{\hat{F}^\dagger}
\newcommand{\oFt}{{\Ftil}}
\newcommand{\oFtdg}{{\Ftil}^\dagger}
\newcommand{\og}{\hat{g}}
\newcommand{\ogdg}{\hat{g}^\dagger}

\newcommand{\oH}{\hat{H}}

\newcommand{\oJ}{\hat{J}}
\newcommand{\oJv}{\hat{\Jv}}

\newcommand{\oK}{\hat{K}}

\newcommand{\omt}{\hat{\mtil}}

\newcommand{\on}{\hat{n}}
\newcommand{\ont}{\hat{\ntil}}
\newcommand{\oN}{\hat{N}}

\newcommand{\ophi}{\hat{\phi}}
\newcommand{\ophidg}{{\hat{\phi}}^\dagger}

\newcommand{\oPhi}{\hat{\Phi}}

\newcommand{\opsi}{\hat{\psi}}
\newcommand{\opsidg}{{\hat{\psi}}^\dagger}
\newcommand{\opsit}{{\psitil}}
\newcommand{\opsitdg}{{{\psitil}}^\dagger}

\newcommand{\cL}{{\cal L}}
\newcommand{\cM}{{\cal M}}

\newcommand{\Btil}{\widetilde B}

\newcommand{\Ftil}{\tilde F}

\newcommand{\ntil}{\tilde n}
\newcommand{\mtil}{\tilde m}

\newcommand{\Ometil}{\widetilde\Ome}

\newcommand{\psitil}{\tilde\psi}

\newcommand{\dv}{\vec{d}}

\newcommand{\cEv}{\vecgr{\cal E}}

\newcommand{\Jv}{\vec{J}}

\newcommand{\kv}{\vec{k}}

\newcommand{\nabv}{\vecgr{\nabla}}

\newcommand{\rv}{\vec{r}}

\newcommand{\iKvar}[1]{{\sssty (#1)}}

\begin{document}

\title{Limitations of squeezing due to collisional decoherence in Bose-Einstein condensates}
\author{T.~Gasenzer\thanks{email: t.gasenzer1@physics.ox.ac.uk}
}
\address{Clarendon Laboratory, Department of Physics, University of Oxford,
Oxford OX1~3PU, United Kingdom\\(\today)}
\maketitle

\begin{abstract} 
We study the limitations for entanglement due to collisional decoherence in a Bose-Einstein condensate.
Specifically we consider relative number squeezing between photons and atoms coupled out from a homogeneous condensate. 
We study the decay of excited quasiparticle modes due to collisions, in condensates of atoms with one or two internal degrees of freedom. 
The time evolution of these modes is determined in the linear response approximation to the deviation from equilibrium.  
We use Heisenberg-Langevin equations to derive equations of motion for the densities and higher correlation functions which determine the squeezing. 
In this way we can show that decoherence due to quasiparticle interactions imposes an important limit on the degree of number squeezing which may be achieved.
Our results are also relevant for the determination of decoherence times in other experiments based on entanglement, e.g.~the slowing and stopping of light in condensed atomic gases using dark states.    
\\[3pt]
\end{abstract}

\setcounter{equation}{0}
\section{Introduction}
Squeezing and entanglement in coherent many particle systems have become important issues in experiments with Bose-Einstein condensates (BECs) as well as in the theoretical work related to them.
Achieving squeezed and/or entangled states of a large number of ultracold atoms would have an enormous impact on the research field which includes quantum information technology (QIT) \cite{QIT}, the development of new time standards \cite{Huelga97} and the test of quantum state reduction \cite{Penrose98}.
In particular, whereas techniques for entangling photons \cite{EntPhotons} are already available for the transmission of quantum information \cite{QCryptQTelep}, schemes for the local storage and processing are still missing.
A promising step in the desired direction represent the experiments which allow for photon pulses being slowed and even stored for a considerably long time in cold and ultracold, condensed atomic gases \cite{Lightslowingstopping} (for a recent review cf.~\cite{Lukin01}).
These experiments are based on electromagnetically induced transparancy \cite{EIT} and use the possibility of dark states, which for the case of many-particle systems involving a condensate may be described as entangled states of atoms and photons.

In \cite{GRB01} we have studied the limitations of relative number squeezing in entangled states of atoms and photons produced by Bragg-scattering of light off a condensate.
Entanglement between photons and atoms has been proposed and studied by different authors before \cite{Ruostekoski98Moore9900}. 
We took into account interactions of the condensate atoms with non-condensed atoms to the degree that the quasiparticles resulting from the diagonalization of the quadratic part of the Hamiltonian remained free.
We have shown that the amount of squeezing in such systems underlies important limits imposed by the collisional interactions of the atoms.
In this paper we would like to extend this work and study the effect of quasiparticle interactions which lead to collisional damping of excitations.
We will find additional constraints on the squeezing considered in \cite{GRB01}.

Moreover, we will consider the case where the non-equilibrium matter wave excitations consist of atoms which are in a different internal state than the condensate atoms.
This is e.g.~relevant in the case of a dark state involving a condensate, a classical coupling laser and quantized probe and atomic modes, which allows to slow and stop light pulses in atomic matter \cite{RGB02}.
The quantized atomic mode is in general in a different internal state than the classical (condensate) mode and the decoherence time is governed by the spin-conserving collisions between the atoms of the different modes.
In this paper we derive the collisional damping rate of an excited atomic mode for such a system with two internal degrees of freedom, in an analogous way as in the single level case.
In \cite{RGB02} we have studied the limitations due to this collisional damping of the light delay and storage times in experiments based on electromagnetically induced transparency.

Mean field theory and its extensions have been widely used to describe stationary properties of homogeneous and trapped dilute, weakly interacting Bose-Einstein condensates at finite temperatures \cite{Dalfovo99,Griffin96,HFB}.
In these approaches, the non-condensate component is treated as a static thermal bath and its dynamical coupling to the condensate oscillations is neglected.
This implies that damping of out-of-equilibrium-excitations is not accounted for.
Time dependent mean field approaches have been introduced \cite{TDHFB,Giorgini98,Rusch99,Giorgini00} to describe the damping in trapped condensates at finite temperatures \cite{Giorgini98}.
The calculated damping rates agree with the results from perturbation theory \cite{Pitaevskii97,Fedichev98}. 
In \cite{Dampingrates,Graham00} damping rates in trapped condensates at finite temperature have been calculated explicitly.
In order to describe the dynamics of a decaying excitation consistently at large times it is necessary to take into account the connection between dissipation and fluctuations. 
To this end it is convenient to use the Heisenberg-Langevin approach  (cf.~e.g.~\cite{MandelWolf}) to treat the time development of the system \cite{Graham98,Graham00}.
In this paper we would like to recall this procedure as applied to the dissipation dynamics of condensates with atoms in a single internal state and extend it to the treatment of the two internal levels case.
Moreover, the Heisenberg-Langevin equations will enable us to derive the time evolution of expectation values of products of two and more quasiparticle operators which we need to study the squeezing described above.
\\ 
 
In the basic, quadratic approximation to the Hamiltonian of the system, including a quartic collisional contact interaction, mean field theory yields the ground state wave function and the spectrum of elementary excitations, which may be understood as non-interacting quasiparticles. 
In this theory, the Heisenberg equations of motion of the Fock operators assume the form
\be
\label{eq0a}
  \dot{\obet}_i(t) = -i\ome_i\obet_i(t),
\ee
where $\obet_i(t)$ is the annihilation operator of the quasiparticle mode labelled by $i$, which satisfies Bose equal-time commutation relations with its hermitian conjugate, $[\obet_i(t),\obetdg_j(t)]=\del_{ij}$. 
$\ome_i$ is the corresponding eigenfrequency. 

Beyond this free field theory the atomic collisions induce interactions of the quasiparticles.
The purpose of this paper is to consider the impact of these interactions on the dynamic behaviour of a BEC system which is initially in a state out of thermal equilibrium. 
Specifically we study the decay behaviour of a single quasiparticle mode whose initial population is different from the equilibrium one. 
As this quasiparticle mode is interacting with the bath of other modes the system will evolve back to a stationary equilibrium state. 
The dynamics of such processes may be described using a time-dependent mean field approach based on the finite temperature extension of the Gross-Pitaevski equation \cite{Giorgini98,Giorgini00}. 
This allows to describe oscillations and damping of the condensate mean field which initially is assumed to be slightly displaced from its stationary form.

As an alternative to the mean-field approach one may adhere for the first to operators and find the equations of motion for the quasiparticle operators. 
In the basis of the non-interacting quasiparticles of the quadratic theory these will be in general coupled non-linear equations. 
For the physical situations we are considering here, we are interested in approximations which allow to write the equation for the excited mode $q$ in the form      
\be
\label{eq0b}
  \dot{\obet}_q(t) = -i(\ome_q+\del\ome_q-\ih\gam_q)\obet_q(t),
\ee
where $\del\ome_q$ is an energy shift and $\gam_q$ a decay width. 
As was shown using the extended time dependent mean-field theory \cite{Giorgini98,Giorgini00}, the decay width arises from the coupling of the excited mode to the bath of the other modes which are assumed to be in thermal equilibrium and to have a broad range of suffiently closely spaced frequencies. 
Eqn.~(\ref{eq0b}) may be used e.g.~to calculate the time dependence of the mean population of mode $q$, which results as an exponential decay with rate $\gam_q$.
It is also clear, however, that the operator which evolves in time as described by (\ref{eq0b}) does not obey the equal time commutation relation, but commutes with $\obetdg_q(t)$ for $t\to\infty$. 
Moreover, according to (\ref{eq0b}) the population of the mode $q$ decays to zero, i.e.~not to the correct equilibrium value at finite temperatures. 
These problems are absent in the mean-field approach, where only the deviation from the stationary condensate field decays and where no operator ordering problems occur.

It is well known in quantum optics that these problems are resolved by the Heisenberg-Langevin theory of dissipative dynamics.
Let us briefly summarize its results.
The equation of motion (\ref{eq0b}) for $\obet_q$ contains a further term.
In the Markov approximation, the equation may be written in the form
\be
\label{eq35}
  \dot{\obet_q}(t)
  = -i(\ome_q+\del\ome_q-\ih\gamma_q)\,\obet_q(t)-\oF_q(t),
\ee
where $\oF_q$ is a Langevin quantum noise operator which has a zero mean value.
The Langevin operator in the Markov approximation obeys the commutation relation
\be
\label{eq37}
  \lk[\oF_q(t),\oFdg_q(t')\rk] = \gamma_q\,\del(t-t').
\ee
This ensures that the equal time commutation relation for $\obet_q(t)$ is conserved in time,
\be
\label{eq38}
   [\obet_q(t),\obetdg_q(t)] = 1.
\ee
In the Markov approximation the noise operator is delta-correlated in time, i.e.~the two-time correlation function is given by
\be
\label{eq39}
   \la\oFdg_q(t)\oF_q(t')\ra = \gam_q n^0_q \del(t-t'),
\ee
where $n^0_q$ is the equilibrium value of the mode population.
Eqn.~(\ref{eq39}) means, that the random forces described by $\oF_q$ have a short memory. 
The equation of motion for the quasiparticle number operator $\on_q=\obetdg_q\obet_q$ becomes
\be
\label{eq40}
   \dot{\on_q}(t) = -\gam_q\,\lk[\on_q(t)-n^0_q\rk] - \oF_q'(t),
\ee
where $\oF_q'$ is the quantum noise operator defined by
\be
\label{eq41}
   \oF_q' 
   \equiv \obetdg_q\oF_q+\oFdg_q\obet_q-\la\obetdg_q\oF_q+\oFdg_q\obet_q\ra,
\ee
which by definition has a zero mean value.
To prove eqn.~(\ref{eq40}), it has to be shown expicitly that $\la\obetdg_q\oF_q+\oFdg_q\obet_q\ra=n^0_q$.
Eqn.~(\ref{eq40}) shows that the quasiparticle population of mode $q$ decays, as expected, exponentially to the equilibrium value $n^0_q$.
Before the population can decay to zero the fluctuations introduced by the coupling to the bath modes force it to settle at the thermal equilibrium value. 


Our paper is organized as follows:
In section \ref{sec2} we review the Heisenberg-Langevin approach to quasiparticle damping for the single internal level case. We use a linear response treatment of the operator time evolution analogous to the one used in \cite{Giorgini98,Giorgini00} for the mean field fluctuation and derive both (\ref{eq37}) and (\ref{eq39}) in the Markov approximation.
In section \ref{sec3} we extend the approach to the case of two internal levels and derive the Beliaev- and Landau-type decay widths as well as the corresponding noise operators.
In section \ref{sec4} we study time evolution equations of the form (\ref{eq40}) for normal ordered products of two and more quasiparticle operators. 
We introduce a quantum regression theorem which allows to use the basic equation (\ref{eq0b}) for the operator to calculate the time evolution of the operator products. 
We use this to calculate the implications of quasiparticle damping for the time evolution of the atom-photon squeezing produced by Bragg scattering of photons off condensates as described in \cite{GRB01}.
Section \ref{sec5} contains our conclusions.

\section{Heisenberg-Langevin equation for systems of atoms in a single internal state}
\label{sec2}
The Hamiltonian of the interacting system of non-relativistic bosonic atoms in the presence of an external trapping potential $V(\rv)$, in the grand canonical description, has the form
\bea
\label{eq1}
  \oK\equiv\oH-\mu\oN 
  &=& \intdd{r}\lk\{ \opsidg(\rv,t)\lk(
  -\frac{\hbar^2\nabv^2}{2m}+V(\rv)-\mu\rk)\opsi(\rv,t)
  +\frac{{g}}{2}\,\opsidg(\rv,t)\opsidg(\rv,t)\opsi(\rv,t)\opsi(\rv,t)\rk\}.
\eea
We have chosen the contact potential approximation for the interatomic potential, $U(\rv)={g}\del(\rv)$, where ${g}$ is given in terms of the s-wave scattering length by ${g}=4\pi\hbar^2a/m$. 
The field operators $\opsi(\rv,t)$ and $\opsidg(\rv,t)$ obey the usual equal-time commutation relations. 
Using these, the equation of motion for the field operator follows to be 
\bea
\label{eq2}
  i\hbar\ptdnd{t}\opsi(\rv,t)
  &=& \lk(-\frac{\hbar^2\nabv^2}{2m}+V(\rv)-\mu\rk)\opsi(\rv,t)
  +{g}\,\opsidg(\rv,t)\opsi(\rv,t)\opsi(\rv,t).
\eea
In the usual treatment of the time evolution of a Bose condensate away from, but close to equilibrium a time dependent condensate wave function is defined by
\be
\label{eq3}
  \Phi(\rv,t)=\la\opsi(\rv,t)\ra,
\ee
where the avarage $\la\cdots\ra$ is with respect to the non-equilibrium initial state of the system. 
The field operator is then decomposed into a condensate and a non-condensate component,
\be
\label{eq4}
  \opsi(\rv,t) = \oPhi(\rv,t) + \opsit(\rv,t),
\ee
and the non-condensate part is defined by $\la\opsit(\rv,t)\ra=0$ \cite{fnpsiops}. 
The condensate part is usually taken to be a c-number, but we will keep its operator character as far as its non-equilibrium contribution is concerned:
\be
\label{eq5}
  \oPhi(\rv,t) = \Phi_0(\rv) + \ophi(\rv,t).
\ee
Here $\Phi_0(\rv)=\la\opsi(\rv)\ra_0$ is the stationary expectation value w.r.t.~the equilibrium state of the BEC and assumed to be real. 
In the usual mean field treatment \cite{Giorgini98,Rusch99,Giorgini00} the condensate part $\oPhi$ is replaced by its non-equilibrium expectation value $\Phi=\la\oPhi\ra$, and therefore $\ophi$ by the fluctuation field $\delta\Phi=\la\ophi\ra$, which is assumed to be small compared to $\Phi_0$. 
In our further procedure we will also assume that the non-equilibrium expectation value of $\ophi$ remains small compared to $\Phi_0$, in order to justify an analogous linear response approach to the time-evolution of the field operator $\ophi$.

We now define the fluctuation operators of the normal and anomalous densities as
\bea
\label{eq6}
  \del\ont &=& \opsitdg\opsit - \ntil^0, \\
\label{eq7}
  \del\omt &=& \opsit\opsit - \mtil^0,
\eea
where we have, as we will do in the following, suppressed the space and time arguments of the operators and densities. 
The non-equilibrium expectation values of the operators (\ref{eq6},\ref{eq7}) are the fluctuation functions $\del\ntil$, $\del\mtil$ defined in \cite{Giorgini98}, which describe the deviation from the equilibrium expectation values $\ntil^0=\la\opsitdg\opsit\ra_0$, $\mtil^0=\la\opsit\opsit\ra_0$.

Inserting eqns.~(\ref{eq4})--(\ref{eq7}) into (\ref{eq2}) we obtain, to linear order in $\ophi$, $\opsit$:
\bea
\label{eq8}
  i\hbar\ptdnd{t}\lk(\opsit+\ophi\rk)
  &=& \lk(\oK_0+2{g}\lk[n_0+\ntil^0\rk]\rk)\,\opsit
      +{g}\lk[n_0+\mtil^0\rk]\opsitdg
  \nonumber\\
  &&\ +\ \lk(\oK_0+2{g}\lk[n_0+\ntil^0
             +\Phi_0\opsitdg+\Phi_0\opsit\rk]\rk)\,\ophi
      +{g}\ophidg\lk[n_0+\mtil^0+2\Phi_0\opsit\rk]\,
  \nonumber\\
  &&\ +\ {g}\Phi_0\lk[2\del\ont+\del\omt\rk],
\eea
where we have defined $\oK_0=-\hbar^2\nabv^2/(2m)+V-\mu$, and where $n_0=|\Phi_0|^2$ is the condensate density. 
In deriving (\ref{eq8}) we have also used that the equilibrium condensate wave function $\Phi_0$ satisfies the generalized stationary Gross-Pitaevskii equation
\be
\label{eq9}
  \lk(\oK_0+{g}\lk[n_0+2\ntil^0+\mtil^0\rk]\rk)\,\Phi_0=0,
\ee
and we have neglected the fluctuation operators 
\be
\label{eq9a}
  \del(\opsitdg\opsit\opsit)
  \equiv\opsitdg\opsit\opsit-2\ntil^0\opsit-\mtil^0\opsitdg.
\ee

In the case, where the system is in equilibrium, the r.h.s.~(\ref{eq8}) reduces to its first line. 
Then, eqn.~(\ref{eq8}) may be solved by introducing the Bogoliubov expansion 
\be
\label{eq10}
  \opsit(\rv,t) = \sum_i\lk[u_i(\rv)\obet_i(t)+v^*_i(\rv)\obetdg_i(t)\rk]
\ee
of the field operator in terms of quasiparticle mode operators. 
In equilibrium, (\ref{eq8}) is linear in the field operators, one has $\obet_i(t)=\obet_i(t_0)\exp(-i\ome_i (t-t_0))$ and the mode functions $u_i$, $v_i$ are the solutions of the Bogoliubov-de Gennes equations \cite{deGennes66}
\be
\label{eq11}
  \lk(\ba{cc}\cL(\rv)&\cM(\rv)\\
            \-\cM^*(\rv)&-\cL^*(\rv)\ea\rk)\lk(\ba{c}u_i(\rv)\\v_i(\rv)\ea\rk) 
  = \hbar\ome_i\,\lk(\ba{c}u_i(\rv)\\v_i(\rv)\ea\rk),
\ee
where $\cL(\rv)=\oK_0(\rv)+2{g}[n_0(\rv)+\ntil^0(\rv)]$, $\cM(\rv)={g}[n_0(\rv)+\mtil^0(\rv)]$. 
A normalization of the mode functions $u_i$, $v_i$ by
\be
\label{eq12}
  \intdd{r}\lk[u^*_i(\rv)u_j(\rv)-v^*_i(\rv)v_j(\rv)\rk] = \del_{ij}
\ee
ensures that the quasiparticle operators $\obet_i$, $\obetdg_i$ satisfy bosonic commutation relations.  

Using (\ref{eq10}) we can expand the fluctuation operators (\ref{eq6}), (\ref{eq7}) in terms of the operators
\bea
\label{eq13}
  \of_{ij} &=& \obetdg_i\obet_j - \del_{ij}n^0_i,\\
\label{eq14}
  \og_{ij} &=& \obet_i\obet_j
\eea
(the notation using $f$ and $g$ follows \cite{Giorgini98}), and find
\bea
\label{eq15}
  \del\ont
  &=& \sum_{ij}\lk\{\lk[u_i^*u_j+v_i^*v_j\rk]\of_{ij}
      +u_iv_j\,\og_{ij}+u^*_iv^*_j\,\ogdg_{ij}\rk\},\\
\label{eq16}
  \del\omt
  &=& \sum_{ij}\lk\{2v_i^*u_j\of_{ij}
      +u_iu_j\,\og_{ij}+v^*_iv^*_j\,\ogdg_{ij}\rk\}.
\eea
We now proceed with solving eqn.~(\ref{eq8}) in the non-equilibrium case.
For simplicity, we assume that the initial state is, in the basis of the quasiparticle modes defined by (\ref{eq10}), a quasiparticle number state.
We assume that a single quasiparticle mode is populated much stronger than the thermal equilibrium population, $n_q(t=0)=\la\obetdg_q(0)\obet_q(0)\ra\gg n^0_i=\la\obetdg_i\obet_i\ra_0=[\exp(\hbar\ome_i/k_BT)-1]^{-1}$ ($i\not=q$). 

We write the fluctuation operator $\ophi$ in the form
\be
\label{eq17}
  \ophi(\rv,t) = u_q(\rv)\obet_q(t)+v^*_q(\rv)\obetdg_q(t).
\ee
In order to justify the linear response treatment, the particle population is still assumed to be much lower than that of the condensate mode, $\la\ophidg(\rv,0)\ophi(\rv,0)\ra=\intdd{r}([|u_q(\rv)|^2+|v_q(\rv)|^2]n_q(t)+|v_q(\rv)|^2)\ll n_0$.

Note that in (\ref{eq17}) we assume $\opsi$ to be defined by (\ref{eq10}), but with the sum over $i$ excluding $q$. Following (\ref{eq6}) and (\ref{eq7}) the sums over $i$ and $j$ in the expansions (\ref{eq15}) and (\ref{eq16}) then also exclude $q$. 
However, we will include the terms proportional to $\opsit\ophi$, $\opsitdg\ophi$ and $\ophidg\opsit$ in (\ref{eq8}) into the term ${g}\Phi_0[2\del\ont+\del\omt]$ which means that in (\ref{eq15}), (\ref{eq16}) the sums over $i$ and $j$ also include $q$, except the combination $i=j=q$, which is neglected in linear response.

Now, inserting (\ref{eq10}) and (\ref{eq17}) in (\ref{eq8}) and using (\ref{eq11}) and (\ref{eq12}), we find the equation of motion for $\obet_q(t)$: 
\bea
\label{eq18}
  \dot{\obet_q}(t)
  &=& -i\ome_q\obet_q(t) -\iohbar {g}\sum_{ij}\Big\{
      2A_{q,ij}^*\,\of_{ij}(t)
      +B_{q,ij}^*\,\og_{ij}(t)+\Btil_{q,ij}^*\,\ogdg_{ij}(t)\Big\},
\eea
where
\bea
\label{eq32}
   A_{q,ij}
   &=& \intdd{r}\Phi_0\lk[
        u_qA_{ij}^*+v_qA_{ji}\rk],
   \nonumber\\
\label{eq33}
   B_{q,ij}
   &=& \intdd{r}\Phi_0\lk[
        u_qB_{ij}^*+v_qC_{ij}^*\rk],
   \nonumber\\
\label{eq34}
   \Btil_{q,ij}
   &=& \intdd{r}\Phi_0\lk[
        u_qC_{ij}+v_qB_{ij}\rk].
\eea
and
\bea
\label{eq19}
   A_{ij} &=& u_i^*u_j+v_i^*v_j+v_i^*u_j,\nonumber\\
\label{eq20}
   B_{ij} &=& u_iu_j+u_iv_j+v_iu_j,\nonumber\\
\label{eq21}
   C_{ij} &=& v_iv_j+u_iv_j+v_iu_j.
\eea
The time evolution equations for the density and pair operators $\of_{ij}$, $\og_{ij}$ may be derived in an analogous way as in \cite{Giorgini98}. 
Inserting (\ref{eq4}) and (\ref{eq5}) into (\ref{eq1}), using (\ref{eq6}), (\ref{eq7}), (\ref{eq9a}), (\ref{eq10}), and (\ref{eq11}), and keeping only terms up to linear order in the fluctuation operators, one finds:
\bea
\label{eq22}
  \oK &=& \sum_{i}\,\hbar\ome_i\obetdg_i\obet_i
  \nonumber\\
  &&\ +\ {g}\intdd{r}\Phi_0\lk[
        2\ophidg\opsitdg\opsit+\ophidg\opsit\opsit+\mathrm{h.c.}\rk]
  \nonumber\\
  &&\ +\ {g}\intdd{r}\lk[
        4\,\del\ntil\,\opsitdg\opsit
        +\del\mtil\,\opsitdg\opsitdg+\del\mtil^*\,\opsit\opsit\rk]
  \nonumber\\
  &&\ +\ {g}\intdd{r}\lk[
        \del(\opsitdg\opsitdg\opsit\opsit)
       +\Phi_0\lk(\del(\opsitdg\opsitdg\opsit)
                 +\del(\opsitdg\opsit\opsit)\rk)\rk] 
  \nonumber\\
  &&\ -\ {g}\intdd{r}\Phi_0\lk[
	(2\ntil^0+\mtil^{0*})\ophi+\mathrm{h.c.}\rk],        
\eea
where
\bea
\label{eq23}
  \del(\opsitdg\opsitdg\opsit\opsit)
  &=& \opsitdg\opsitdg\opsit\opsit
     -4\la\opsitdg\opsit\ra\,\opsitdg\opsit
     -\la\opsitdg\opsitdg\ra\,\opsit\opsit
     -\la\opsit\opsit\ra\,\opsitdg\opsitdg,
  \\
\label{eq24}
  \del(\opsitdg\opsitdg\opsit)
  &=& \opsitdg\opsitdg\opsit-2\ntil^0\opsitdg-\mtil^{0*}\opsit, 
  \\
\label{eq25}
  \del\ntil
  &=& \la\opsitdg\opsit\ra-\ntil^0=\la\del\ont\ra, 
  \\
\label{eq26}
  \del\mtil
  &=& \la\opsit\opsit\ra-\mtil^0=\la\del\omt\ra.        
\eea
Then, using (\ref{eq10}), (\ref{eq13}), (\ref{eq14}) the equations of motion for $\of_{ij}$ and $\og_{ij}$ read ($i,j\not=q$)
\bea
\label{eq27}
  \ptdnd{t}\of_{ij}
  &=& \iohbar\lk[\oK,\obetdg_i\obet_j\rk]
  \nonumber\\
  &=& -i(\ome_j-\ome_i)\of_{ij}-2\iohbar {g}(n^0_i-n^0_j)\intdd{r}\Phi_0\lk[
      A_{ij}^*\ophi+A_{ji}\ophidg\rk],
  \\
\label{eq28}
  \ptdnd{t}\og_{ij}
  &=& \iohbar\lk[\oK,\obet_i\obet_j\rk]
  \nonumber\\
  &=& -i(\ome_j+\ome_i)\og_{ij}-2\iohbar {g}(1+n^0_i+n^0_j)\intdd{r}\Phi_0\lk[
      B_{ij}^*\ophi+C_{ij}^*\ophidg\rk],
\eea
In deriving the time evolution (\ref{eq27},\ref{eq28}) of the fluctuation operators from (\ref{eq22}), the terms proportional $\del\ntil$ and $\del\mtil$ which describe the coupling to the fluctuations of the bath modes have been neglected. 
Also the terms arising from the fluctuations $\del(\opsitdg\opsitdg\opsitdg\opsit)$, $\del(\opsitdg\opsitdg\opsit)$ have not been included. 
Note also, that in the cases $i=q$ or $j=q$, for which the differential equations (\ref{eq27}) and (\ref{eq28}) take the same form, further terms in $\oK$ have to be taken into account. 
These terms are of second order in $\ophi$, $\ophidg$ and have been neglected in (\ref{eq22}). We will in the following assume these special cases to be included.
Equations (\ref{eq27}) and (\ref{eq28}) are then the analogues of eqns.~(26) and (27) of \cite{Giorgini98}.

We can integrate (\ref{eq27}) and (\ref{eq28}) formally and insert the solutions back into (\ref{eq18}).
\bea
\label{eq28a}
  \dot{\obet_q}(t)
  &=& -i\ome_q\obet_q(t)
  \nonumber\\
  &&\ -\iohbar {g}\sum_{ij}\Big\{
      2A_{q,ij}^*\,e^{-i(\ome_j-\ome_i)t}\of_{ij}(0)
      +B_{q,ij}^*\,e^{-i(\ome_j+\ome_i)t}\og_{ij}(0)
      +\Btil_{q,ij}^*\,e^{i(\ome_j+\ome_i)t}\ogdg_{ij}(0)\Big\}
  \nonumber\\
  &&\ -2({g}/\hbar)^2\sum_{ij}\,\int dt\,\Big\{
       2(n^0_i-n^0_j)\,|A_{q,ij}|^2\,e^{i(\ome_i-\ome_j)(t-t')}
       \nonumber\\
   && \phantom{2({g}/\hbar)^2\int dt\,}+\
       (1+n^0_i+n^0_j)\,\Big[
            |B_{q,ij}|^2\,e^{-i(\ome_i+\ome_j)(t-t')}
        -|\Btil_{q,ij}|^2\,e^{i(\ome_i+\ome_j)(t-t')}\Big]
	\Big\}.
\eea
The terms involving the time integrals then lead, in the Markov approximation, to an energy shift $\del\ome_q$ and a decay width $\gam_q$. 
The energy shift may be calculated in the Popov approximation ($\mtil^0=0$) \cite{Popov64}, where it vanishes for $q\rar0$, i.e.~leaves the spectrum gapless as required by the Hugenholtz-Pines theorem \cite{Hugenholtz59,Griffin96}. 
For temperatures $k_BT\ll\mu$, however, the anomalous density $\mtil^0$ can not be neglected in the calculations.
A self consistent solution of the Hartree-Fock-Bogoliubov equations including $\mtil^0$, together with a renormalization of the coupling ${g}$ results in a gapless excitation spectrum \cite{Morgan00}.

We are here primarily interested in the damping of the excited quasiparticle mode. 
The decay width $\gam_q$ follows, analogously to \cite{Giorgini98} to be
\be
\label{eq29}
  \gam_q = \gam_L(q) + \gam_B(q),
\ee
$\gam_L$ and $\gam_B$ are known as the Landau and Beliaev decay widths \cite{Giorgini98,Pitaevskii97,Fedichev98,Beliaev58} and are found from (\ref{eq28a}) to be:
\bea
\label{eq30}
   \gam_L(q)
   &=& 8\pi ({g}/\hbar)^2\sum_{ij}\,|A_{q,ij}|^2(n^0_i-n^0_j)\,
       \del(\ome_q+\ome_i-\ome_j),
   \\
\label{eq31}
   \gam_B(q)
   &=& 4\pi ({g}/\hbar)^2\sum_{ij}\,|B_{q,ij}|^2(1+n^0_i+n^0_j)\,
       \del(\ome_q-\ome_i-\ome_j).
\eea
To derive (\ref{eq30},\ref{eq31}) we have chosen the Markov approximation.
This means that we assume that the sum over $i$ and $j$ over the appearing products of coupling functions (\ref{eq32}), multiplied by the energy exponentials, give functions of $\tau=t-t'$ which are sharply peaked around $\tau=0$.
The conditions for this assumption to hold are generally that the energies of the bath modes are closely spaced compared to the energy of the decaying mode and that the couplings are sufficiently well behaved at large energies.
The time dependence of the sharply peaked function may then be approximated as Markovian, i.e.~by a delta-distribution, and integration over time yields the energy conservation delta-distributions in (\ref{eq30},\ref{eq31}).


The first term on the r.h.s.~of (\ref{eq28a}) is the Langevin noise operator corresponding to the decay of the quasiparticle mode. 
Hence, the equation of motion for $\obet_q$ may be written in the form (\ref{eq35}), where the decay width $\gamma_q$ is given by (\ref{eq29})--(\ref{eq31}) and the Langevin noise operator is 
\bea
\label{eq36}
  \oF_q(t)
  &=& \iohbar {g}\sum_{ij}\,\Big\{
      2A_{q,ij}^*e^{-i(\ome_j-\ome_i)t}\of_{ij}(0)
      \nonumber\\
  &&\phantom{\iohbar {g}\sum_{ij}}+\
      B_{q,ij}^*e^{-i(\ome_j+\ome_i)t}\og_{ij}(0)
      \nonumber\\
  &&\phantom{\iohbar {g}\sum_{ij}}+\
      \Btil_{q,ij}^*e^{i(\ome_j+\ome_i)t}\ogdg_{ij}(0)\Big\}
\eea
Using this expression we will now show that in the Markov approximation $\oF_q$ obeys the commutation relation (\ref{eq37}) and its two-time correlation function is given by (\ref{eq39}). 

To this end we need the commutation relations 
\bea
\label{eq42}
   \lk[\of_{ij}(0),\ofdg_{kl}(0)\rk] 
   &=& \del_{ik}\del_{jl} (n^0_i-n^0_j),
   \\
\label{eq43}
   \lk[\og_{ij}(0),\ogdg_{kl}(0)\rk] 
   &=& \lk(\del_{il}\del_{jk}+\del_{ik}\del_{jl}\rk) (1+n^0_i+n^0_j).
\eea
In deriving these we have neglected on the right hand sides terms proportional to the fluctuation operators $\of_{ij}$. 
Using (\ref{eq42},\ref{eq43}) and the fact, that mixed commutators of $\of_{ij}$ and $\og_{kl}$ only yield fluctation operators $\og_{ij}$ we arrive at
\bea
\label{eq44}
   \lk[\oF_q(t)e^{i\ome_qt},\oFdg_q(t')e^{-i\ome_qt'}\rk] 
   &=& 2({g}/\hbar)^2\sum_{ij}\,\Big\{
       2(n^0_i-n^0_j)\,|A_{q,ij}|^2\,e^{i(\ome_q+\ome_i-\ome_j)(t-t')}
       \nonumber\\
   && \phantom{2({g}/\hbar)}+\
       (1+n^0_i+n^0_j)\,\Big[
            |B_{q,ij}|^2\,e^{i(\ome_q-\ome_i-\ome_j)(t-t')}
       \nonumber\\
   && \phantom{2({g}/\hbar)^2\,(1+n^0_i+n^0_j)}-\
        |\Btil_{q,ij}|^2\,e^{i(\ome_q+\ome_i+\ome_j)(t-t')}\Big]
	\Big\}.
\eea
In the Markov approximation the sum over $i,j$ is replaced by a function proportional to $\del(t-t')$. 
Integration of (\ref{eq44}) and comparison with (\ref{eq29})--(\ref{eq31}) then shows that the coefficient of this delta-distribution is given by $\gam_q$, as stated in (\ref{eq37}).

Eqn.~(\ref{eq39}) may be proven in a similar way, using
\bea
\label{eq45}
   \la\ofdg_{ij}(0)\of_{kl}(0)\ra 
   &=& \del_{ik}\del_{jl} n^0_j(n^0_i+1),
   \\
\label{eq46}
   \la\ogdg_{ij}(0)\og_{kl}(0)\ra 
   &=& \lk(\del_{il}\del_{jk}+\del_{ik}\del_{jl}\rk) n^0_in^0_j.
\eea
In deriving these relations we have made use of Wick's theorem. 
Taking into account still the respective energy matchings in the different terms, the population factors become
\bea
\label{eq47}
   \del(\ome_q+\ome_i-\ome_j)\, n^0_j(n^0_i+1) 
     &=& \del(\ome_q+\ome_i-\ome_j)\, n^0_q(n^0_i-n^0_j),
   \\
\label{eq48}
   \del(\ome_q-\ome_i-\ome_j)\, n^0_in^0_j
     &=& \del(\ome_q-\ome_i-\ome_j)\,  n^0_q(1+n^0_i+n^0_j).
\eea
where we used the Planck distribution $n^0_i=\la\obetdg_i\obet_i\ra_0=[\exp(\hbar\ome_i/k_BT)-1]^{-1}$. 
Together with (\ref{eq43}) we arrive at
\bea
\label{eq49}
   \la\oFdg_q(t)\oF_q(t')\ra e^{-i\ome_q(t-t')} 
   &=& n^0_q\lk[\oF_q(t),\oFdg_q(t')\rk]\,e^{-i\ome_q(t-t')}.
\eea
where the r.h.s.~is given by (\ref{eq44}). 
This yields, in the Markov approximation, the identity (\ref{eq39}) for the correlation function. 

\section{Heisenberg-Langevin equation for systems of atoms in two internal states}
\label{sec3}
In the following we would like to extend the procedure reviewed in sec.~\ref{sec2} to the case with two internal atomic degrees of freedom, where the atoms in the excited mode $q$ are in a different internal state than the condensate atoms.
Such situations are important e.g.~in the context of the dynamics of dark states involving Bose-Einstein condensates which are the basis for experiments using electromagnetically induced transparency \cite{EIT} for the slowing and stopping of light pulses in atomic matter \cite{Lightslowingstopping}.
The decay of the dark state induced by atomic collisions limits the maximum time for which a light pulse may be stored in collective atomic excitations of condensates \cite{RGB02}.

The interactions which cause the quasiparticle decay will be mainly collisions between atoms in the excited mode and condensate atoms, such that they do not change the respective internal states. 
Let us call the two internal levels of the atoms $b$ and $c$, the Bose-condensed atoms being in internal state $|\,b\ra$. 
The Hamiltonian in the grand canonical ensemble then has the form
\bea
\label{eq50}
  \oK &\equiv& \oH-\mu\oN = \oK_b + \oK_c + \oK_{bc},
  \\
\label{eq51}
  \oK_b
  &=& \intdd{r}\lk\{ \opsidg_b\oK_{0,b}\opsi_b
  +\frac{g_{bb}}{2}\,\opsidg_b\opsidg_b\opsi_b\opsi_b\rk\},
  \\
\label{eq52}
  \oK_c
  &=& \intdd{r}\lk\{ \opsidg_c\lk(\oK_{0,c}+\ome_{cb}\rk)\opsi_c
  +\frac{g_{cc}}{2}\,\opsidg_c\opsidg_c\opsi_c\opsi_c\rk\},
  \\
\label{eq53}
  \oK_{bc}
  &=& \intdd{r} g_{bc}\,\opsidg_c\opsidg_b\opsi_c\opsi_b,
\eea
where $\oK_{0,\nu}=-\hbar^2\nabv^2/(2m)+V_{\nu}(\rv)-\mu$, $\nu=b,c$, and $\ome_{cb}=\ome_c-\ome_b$ is the frequency difference of the internal levels.
$g_{\mu\nu}$, $\mu,\nu=b,c$ are the collisonal coupling constants for the respective processes being related to the respective scattering lengths by $g_{\mu\nu}=4\pi\hbar^2a_{\mu\nu}/m$.
As for the single level case we split the field operators for the atoms in level $b$ into mean-field and fluctuation parts,
\be
\label{eq54}
  \opsi_b(\rv,t) = \Phi_0(\rv) + \opsit_b(\rv,t),
\ee
and the operator for the atoms in $c$ into fluctuation and excited modes:
\be
\label{eq55}
  \opsi_c(\rv,t) = \opsit_c(\rv,t) + \ophi_c(\rv,t),
\ee
With these, we find the equation of motion for the excitation operator $\ophi_c$ in an analogous way as in the single level case:
\bea
\label{eq56}
  i\hbar\ptdnd{t}\ophi_c
  &=& \lk(\oK_{0,c}+2g_{cc}\ntil^0_c+g_{bc}\lk[n_0+\ntil^0_b\rk]\rk)\,\ophi_c
      +g_{cc}\mtil^0_c\ophidg_c\,
  \nonumber\\
  &&\ +\ g_{bc}\Phi_0\lk[\del\ont_{bc}+\del\omt_{bc}\rk],
\eea
where $\ntil^0_{\nu}=\la\opsitdg_{\nu}\opsit_{\nu}\ra_0$, $\mtil^0_{\nu}=\la\opsit_{\nu}\opsit_{\nu}\ra_0$ are the mean equilibrium density and anomalous density of the non-condensed fraction in $\nu=b,c$. 
The fluctuation operators $\del\ont_{bc}=\opsitdg_b\opsit_c$, $\del\omt_{bc}=\opsit_b\opsit_c$ are defined as in (\ref{eq6},\ref{eq7}), with their equilibrium mean values though being zero. 

In order to derive from (\ref{eq56}) the equation of motion for the Fock operator of the excited mode we use the decomposition (\ref{eq10}) for $\opsit_b$ and a particle mode decomposition for atoms in the internal state $c$:
\bea
\label{eq57}
  \opsit_c(\rv,t) 
  &=& \sum_i\,\psitil_{c,i}(\rv)\oc_i(t).
\eea
Here the mode functions $\psitil_{c,i}$ are solutions of the equation 
\be
\label{eq58}
  \cL^c(\rv)\psitil_{c,i}(\rv) = \hbar\ome^c_i\,\psitil_{c,i}(\rv),
\ee
where $\cL^c(\rv)=\oK_{0,c}(\rv)+\ome_{cb}+2g_{cc}\ntil^0_c+g_{bc}\lk[n_0+\ntil^0_b\rk]$, and where we have chosen the approximation $\cM^c(\rv)=g_{cc}\mtil^0_c=0$. 
Note that the eigenfunctions then are not quasiparticle modes since the zero mode in $c$ is not macroscopically populated. The equation of motion for $\oc_q$ then reads
\bea
\label{eq59}
  \ocd_q(t)
  &=& -i\ome^c_q\oc_q(t) -\iohbar g_{bc}\sum_{ij}\Big\{
       A_{q,ij}^{bc*}\,\of^{bc}_{ij}(t)
      +B_{q,ij}^{bc*}\,\og^{bc}_{ij}(t)\Big\},
\eea
where
\bea
\label{eq60}
   A^{bc}_{q,ij}
   &=& \intdd{r}\Phi_0\lk[
        \psitil_{c,q}(u_i+v_i)\,\psitil_{c,j}^*\rk],
   \\
\label{eq61}
   B^{bc}_{q,ij}
   &=& \intdd{r}\Phi_0\lk[
        \psitil_{c,q}(u_i^*+v_i^*)\,\psitil_{c,j}^*\rk],
\eea
and
\bea
\label{eq62}
  \of_{ij}^{bc} &=& \obetdg_i\oc_j,\\
\label{eq63}
  \og_{ij}^{bc} &=& \obet_i\oc_j.
\eea
In order to solve (\ref{eq59}) we need, as in the single level case, the time evolution of the operators $\of_{ij}^{bc},\og_{ij}^{bc}$. 
To this end we write down the relevant terms of the Hamiltonian $\oK$ which contribute to the commutator with these operators. 
Using (\ref{eq54}) and (\ref{eq55}) as well as (\ref{eq57}) and (\ref{eq58}) they read
\bea
\label{eq64}
  \oK_c+\oK_{bc} 
  &=& \sum_{i}\,\hbar\ome_i\ocdg_i\oc_i
  + g_{cc}\intdd{r}
        \del(\opsitdg_c\opsitdg_c\opsit_c\opsit_c), 
  \nonumber\\
  &&\ +\ g_{bc}\intdd{r}\Phi_0\lk[
        \opsitdg_b\ophidg_c\opsit_c+\ophidg_c\opsit_c\opsit_b+\mathrm{h.c.}\rk]
  \nonumber\\
  &&\ +\ g_{bc}\intdd{r}\lk[
        \del\ntil_b\,\opsitdg_c\opsit_c+\del\ntil_c\,\opsitdg_b\opsit_b+\lk(
	\del\ntil_{bc}\,\opsitdg_c\opsit_c+\del\mtil_{bc}^*\,\opsit_c\opsit_b
	+\mathrm{h.c.}\rk)\rk]
  \nonumber\\
  &&\ +\ g_{bc}\intdd{r}\lk[
        \del(\opsitdg_c\opsitdg_b\opsit_c\opsit_b)
       +\Phi_0\lk(\del(\opsitdg_c\opsit_c\opsit_b)+\mathrm{h.c.}\rk)\rk],
\eea
where 
\bea
\label{eq65}
  \del(\opsitdg_c\opsitdg_b\opsit_c\opsit_b)
  &=& \opsitdg_c\opsitdg_b\opsit_c\opsit_b
     -\la\opsitdg_c\opsit_c\ra\,\opsitdg_b\opsit_b
     -\la\opsitdg_b\opsit_b\ra\,\opsitdg_c\opsit_c
     -\lk(
      \la\opsitdg_c\opsit_b\ra\,\opsitdg_b\opsit_c
     -\la\opsitdg_c\opsitdg_b\ra\,\opsit_c\opsit_b +\mathrm{h.c.}\rk),
  \\
\label{eq66}
  \del(\opsitdg_c\opsitdg_c\opsit_b)
  &=& \opsitdg_c\opsitdg_c\opsit_b-\ntil_c^0\opsit_b,        
  \\
\label{eq67}
  \del\ntil_c
  &=& \la\opsitdg_c\opsit_c\ra, 
\eea
and $\del\ntil_{bc}=\la\del\ont_{bc}\ra$, $\del\mtil_{bc}=\la\del\omt_{bc}\ra$.
The operator $\del(\opsitdg_c\opsitdg_c\opsit_c\opsit_c)$ and the functions $\del\ntil_b$ are defined as in (\ref{eq23}) and (\ref{eq25}) respectively.
From the terms on the right hand side of (\ref{eq64}) only the quadratic one and the ones in the second line will contribute to the leading order time evolution of $\of_{ij}^{bc}$ and $\og_{ij}^{bc}$. 
The equations of motion for these operators are then found to be
\bea
\label{eq68}
  \ptdnd{t}\of_{ij}^{bc}
  &=& -i(\ome^c_j-\ome^b_i)\of_{ij}^{bc}
      -\iohbar g_{bc}(n^b_i-n^c_j)A^{bc}_{q,ij}\oc_q(t),
  \\
\label{eq69}
  \ptdnd{t}\og_{ij}^{bc}
  &=& -i(\ome^c_j+\ome^b_i)\og_{ij}^{bc}
      -\iohbar g_{bc}(n^b_i-n^c_j)B^{bc}_{q,ij}\oc_q(t),
\eea
where we have neglected terms of higher order in the fluctuation opertors and used $\ophi_q=\psitil_{c,q}\oc_q$.
$n_i^b$ and $n_j^c$ are the equilibrium populations of modes $i$ and $j$ in internal states $b$ and $c$, respectively. 
Analogously to the single level case we then obtain, in the Markov approximation, the equation of motion for $\oc_q$ in the form (\ref{eq35}). 
We find that the decay rate $\gam_q=\gam_L(q)+\gam_B(q)$ has Landau- and Beliaev-like contributions (\ref{eq29}) which read
\bea
\label{eq70}
   \gam_L(q)
   &=& 2\pi (g_{bc}/\hbar)^2\sum_{ij}\,|A^{bc}_{q,ij}|^2(n^0_i-n^0_j)\,
       \del(\ome_q+\ome_i-\ome_j),
   \\
\label{eq71}
   \gam_B(q)
   &=& 2\pi (g_{bc}/\hbar)^2\sum_{ij}\,|B^{bc}_{q,ij}|^2(1+n^0_i+n^0_j)\,
       \del(\ome_q-\ome_i-\ome_j),
\eea
The Langevin noise operator is given by
\bea
\label{eq72}
  \oF_q(t)
  &=& \iohbar g_{bc}\sum_{ij}\,\Big\{
      A_{q,ij}^{bc*}e^{-i(\ome^c_j-\ome^b_i)t}\of^{bc}_{ij}(0)
      \nonumber\\
  &&\phantom{\iohbar {g}\sum_{ij}}+\
      B_{q,ij}^{bc*}e^{-i(\ome^c_j+\ome^b_i)t}\og^{bc}_{ij}(0)\Big\}.
\eea
The proofs that (\ref{eq72}) obeys the commutator (\ref{eq37}) and eqn.~(\ref{eq39}) for the two-time correlation function are conducted in analogy to the single mode case, eqns.~(\ref{eq42})--(\ref{eq49}).

Our result (\ref{eq70},\ref{eq71}) for the damping rate $\gam_q=\gam_L(q)+\gam_B(q)$ may e.g.~be used to calculate the loss of coherence in a dark state used to delay and store light pulses.
In \cite{RGB02} we have calculated the damping rate for a homogeneous condensate and found that the Beliaev rate for small momenta $q$ of the decaying mode is proportional to $q^5$ as in the case where the internal atomic state is the same as that of the condensate atoms. The rate differs only by a constant numerical factor: 
$\gamma_C(q)_{T=0} \approx \hbar q^5/96 m \pi n_0$  compared to $\gamma_C(q)_{T=0} \approx 3\hbar q^5/320 m \pi n_0$ in the single level case (assuming $a_{bc}=a_{bb})$.

\section{Dissipative dynamics of higher correlation functions: Atom-photon squeezing}
\label{sec4}
\begin{figure}[b]
\begin{center}
\setlength{\unitlength}{0.8mm}
\begin{picture}(100,48)(0,0)
\thinlines
\put( 0, 5){\line(1,0){40}}
\put(60,10){\line(1,0){40}}
\put(30,45){\line(1,0){38}}
\multiput(30,35)(4,0){10}{\line(1,0){2}}
\put(30, 5){\vector(1,3){10}}
\put(60,35){\vector(1,-3){8.25}}
\put(0,8){{$|\,g;\,k\ra,\ome_k$}}
\put(80,13){{$|\,g;\,k+\Del k,\ome_{k+\Del k}\ra$}}
\put(30,48){{$|\,e\ra$}}
\put(47,39){$\Del$}
\multiput(9.2,20.2)(4,0){6}{\oval(2,2)[t]}
\multiput(8.8,19.8)(4,0){6}{\oval(2,2)[t]}
\multiput(8.8,20.2)(4,0){6}{\oval(2,2)[t]}
\multiput(9.2,19.8)(4,0){6}{\oval(2,2)[t]}
\multiput(9,20)(4,0){6}{\oval(2,2)[t]}
\multiput(11,20)(4,0){6}{\oval(2,2)[b]}
\multiput(11.2,20.2)(4,0){6}{\oval(2,2)[b]}
\multiput(11.2,19.8)(4,0){6}{\oval(2,2)[b]}
\multiput(10.8,20.2)(4,0){6}{\oval(2,2)[b]}
\multiput(10.8,19.8)(4,0){6}{\oval(2,2)[b]}
\put(31.8,20.2){\vector(1,0){3.2}}
\put(31.8,20){\vector(1,0){3.2}}
\put(31.8,19.8){\vector(1,0){3.2}}
\multiput(65,23)(4,0){6}{\oval(2,2)[t]}
\multiput(67,23)(4,0){6}{\oval(2,2)[b]}
\put(87.9,23){\vector(1,0){3}}
\put(8,24){{$k_l,\Ome_l$}}
\put(72,27){{$k_s,\Ome_s$}}
\end{picture}
\caption[]{\label{fig1}\fns Bragg transition between different modes of the center of mass motion of atoms in the internal ground state. The motional structure of the excited state may be neglected.}
\end{center}
\end{figure}
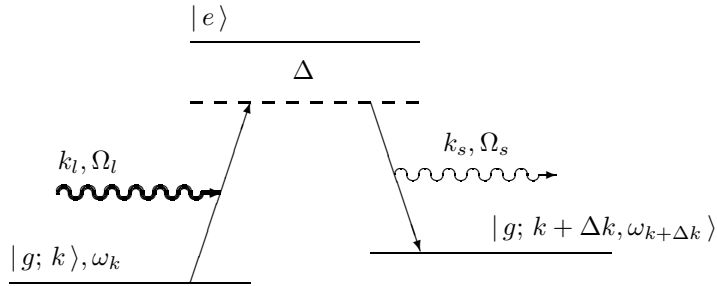

We would now like to apply our formalism in order to calculate the dissipative dynamics of expectation values of two and four quasiparticle operators.
This will enable us to compute the time evolution of the relative number squeezing between atoms and photons produced in photon scattering off condensates.
Such schemes have been proposed in \cite{Ruostekoski98Moore9900}.
In extending this work in \cite{GRB01} we have considered the production of atom-photon squeezing using a finite homogeneous Bose-Einstein condensate of interacting atoms at zero temperature, illuminated by laser light with given frequency $\Ome_l$ detuned slightly from an atomic transition, cf.~fig.\ref{fig1}.
%
%
%
%
%
%

%
The laser photons with momentum $k_l$ predominantly interact with the condensate atoms ($k=0$) and may transfer them in a centre of mass mode with momentum $\Del k$ (see fig.~\ref{fig1}) \cite{VectorNotation}.
Such processes yield a state with relative number squeezing in a particular mode pair, i.e.~the variance of the relative occupation number is suppressed as compared to the classical case.
We computed the squeezing parameter
\be
\label{eq73}
   \xi_3 := \lk[\Del(n_a-n_b)\rk]^2/\la\on_a+\on_b\ra,
\ee
which measures the relative number variance $[\Del(n_a-n_b)]^2=\la (\on_a-\on_b)^2\ra - \la\on_a-\on_b\ra^2$ in relation to its classical limit given by the sum of mean occupations $\la\on_a+\on_b\ra=\la\oadg\oa+\obdg\ob\ra$, of the scattered photon mode and the recoiling atom mode.
(The index '$3$' of $\xi_3$ refers to the analogy of the relative number squeezing with spin squeezing. $\xi_3$ measures the variance of the 3-component $\oJ_3=\eh(\on_a-\on_b)$ compared to half the total value $J/2=\ev\la\on_a+\on_b\ra$ of an abstract angular momentum. This is defined by the Fock operators $\oa$, $\ob$, using the Schwinger representation of angular momentum.
For the issue of spin squeezing and the definition (\ref{eq73}) cf.~\cite{Kitagawa93Wineland94}.
It will be instructive to study also the squeezing of $\oJ_1=\eh(\oadg\ob+\obdg\oa)$ and $\oJ_2=\edzi(\oadg\ob-\obdg\oa)$ for which we define $\xi_i:=[\Del J_i]^2/(J/2)$, $i=1,2$.

In the classical (uncorrelated) limit $\xi_3=1$, whereas for maximum squeezing we have $\xi_3=0$.
There are two reasons why such perfect squeezing is spoilt in an experiment: Collisional atomic interactions and photon (rescattering) processes different from those which produce the squeezed state.
In \cite{GRB01} we focused on the limitations to the maximum squeezing due to those atomic interactions which are taken into account in the basic Hartree-Fock-Bogoliubov approximation, in the Hamiltonian quadratic in the particle operators.
There we have shown that ``unwanted'' collisions amongst the particles present profound limitations to the achievable squeezing, with correlations between atomic modes in the initial state being the main limiting factor at small times.

Here we would like to extend our work to include interactions between quasiparticles described by the third order terms in the Hamiltonian considered in section \ref{sec2}.
The main effect at this order of the approximation is collisional loss from the excited modes which will induce decay of the population of the recoiling atomic mode.
This will also affect the time evolution of the correlation functions $(\Del n_a)^2$, $(\Del n_b)^2$, and  $\la\!\!\la\on_a\on_b\ra\!\!\ra=\la\on_a\on_b\ra-\la\on_a\ra\la\on_b\ra$, which enter the squeezing parameter (\ref{eq73}).
We will neglect energy shifts at this order.
\\

In order to be able to compute the dynamics of the occupation numbers and the four-operator expectation values including dissipation, we will in general have to find equations of motion of the type (\ref{eq40}).
Taking into account the interaction with the laser, the equations for the different expectation values will be coupled.
The task would then be to find the equilibrium expectation values, to which the observables will evolve at large times, as e.g.~$n^0_q$ in the case of eqn.~(\ref{eq40}).
As we are here interested in the zero-temperature case only, we expect these equilibrium limits to be zero in general. 
It should then be justified to take the simpler route and use a regression theorem, i.e.~the basic equations for the Fock operators, like (\ref{eq0b}), in conjunction with Wick's theorem, to calculate the time dependence of the expectation values.
Such a procedure is readily justified in the case without decay.
We will have to show, however, that such a regression theorem may also be used for the dissipative case, and give precise rules for its application.

We start with the dynamic equation for the operators involved in the process to be studied.
In the homogeneous limit, the quasiparticle modes are denoted by their momenta.
We recall that the fluctuation field operator (\ref{eq10}) may be expanded in terms of particle Fock operators, $\opsit(r,t)=V^{-1/2}\sum_k\ob_k\exp(ik\cdot r)$, and the particle operators may be expressed by quasiparticle ones through $\ob_{k}= u_k \obet_k + v_k \obetdg_{-k}$ (cf.~(\ref{eq10})). 
The quasiparticle transformation coefficients are given by
$u_k = (1-\alp_k^2)^{-1/2}$, 
$v_k^2=u_k^2-1$, with
$\alp_k = 1+\kbar^2-\kbar(2+\kbar^2)^{1/2}$,
$\kbar = |k|/k_0$,
$k_0 = \sqrt{8\pi an_0}$, $n_0$ being the condensate density.
The Bogoliubov quasiparticle frequencies are 
$\ome_k = (\hbar k_0^2/2m)\,\kbar\sqrt{2+\kbar^2}$.

We assume two-photon resonance, where the difference $\Del\Ome=\Ome_l-\Ome_s$ of the frequencies of the laser, $\Ome_l$, and of the scattered photon, $\Ome_s$, equals the frequency of the recoiling atom mode, $\Del\Ome=\ome_{\Del k}$, with $\Del k=k_l-k_s$ being the recoil momentum.
If we neglect non-resonant terms, the coupled equations of motion read
\bea
\label{eq74}
  \frac{d}{dt}
  \lk(
  \ba{c}\obettil_{\Del k}(t)\\ \oatdg_s(t)\ea\rk)
  &=& \lk(
    \ba{cc} -\gam_{\Del k}/2 & -i\Ometil \\ i\Ometil & 0\ea\rk)
    \lk(
    \ba{c}\obettil_{\Del k}(t)\\ \oatdg_s(t)\ea\rk)
    -\lk(
    \ba{c}\oFt_{\Del k}(t) \\ 0 \ea\rk),
  \\
\label{eq74b}
  \dot{\obettil}_{\Del k}(t)
  &=& -(\gam_{\Del k}/2)\obettil_{\Del k}(t) - \oFt_{-\Del k}(t).
\eea
Here we have chosen an interaction picture by defining $\obettil_q(t)=\obet_q(t)\exp(i\ome_qt)$, $\oat_s(t)=\oa_s(t)\exp(i\Ome_st)$, $\oFt_q(t)=\oF_q(t)\exp(i\ome_qt)$, and $\Ometil=\Ome(t)\exp(i\Ome_lt)$. 
We denote by $\Ometil = -2{|\dv_{ge}\cdot\cEv^\iKvar{+}(k_1)|\,g_{ge}(k_2)} /{\Del}$ the two photon Rabi frequency, given in terms of the dipole momentum $\dv_{ge}$ between the internal ground and excited states, of the positive frequency part of the electric field describing the laser mode, the coupling $g_{ge}(k_2)$ between the scattered photon mode and the atom, and the detuning $\Del$ from the excited state.
Neglecting the decay width $\gam_{\Del k}$ of the quasiparticle mode as well as the corresponding Langevin operator $\oF_{\Del k}$, we may straightforwardly calculate the time evolution of the particle occupation numbers $\la\on^a_{k_s}\ra$, $\la\on^b_{\pm\Del k}\ra$, and of the squeezing parameter $\xi_3=[(\Del n^a_s)^2+(\Del n^b_{\Del k})^2-2\,\la\!\!\la\on^a_s\on^b_{\Del k}\ra\!\!\ra]/\la\on^a_{s}+\on^b_{\Del k}\ra$, using their expressions in terms of quasiparticle operators given above:
\bea
\label{eq75}
   \la\on^a_{s}\ra
   &=& \la\oa_s\oadg_s\ra-1,
   \\
\label{eq76}
   \la\on^b_{\pm\Del k}\ra
   &=& u_{\Del k}^2\la\obetdg_{\pm\Del k}\obet_{\pm\Del k}\ra
      +v_{\Del k}^2\lk(\la\obetdg_{\mp\Del k}\obet_{\mp\Del k}\ra+1\rk),
   \\
\label{eq77}
   (\Del n^a_s)^2
   &=& \la\on^a_{s}\ra \lk(\la\on^a_{s}\ra+1\rk),
   \\
\label{eq78}
   (\Del n^b_{\Del k})^2
   &=& \la\on^b_{\Del k}\ra\lk(\la\on^b_{\Del k}\ra+1\rk),
   \\
\label{eq79}
   \la\!\!\la\on^a_s\on^b_{\Del k}\ra\!\!\ra
   &=& u_{\Del k}^2|\la\oa_s\obet_{\Del k}\ra|^2.
\eea

To be able to include collisional loss, however, a more careful approach is required.
In appendix \ref{appA} we show that the dynamics of the expectation values appearing on the right hand sides of (\ref{eq75},\ref{eq76},\ref{eq79}) is given by the system of equations 
\bea
\label{eq82}
  \lk(
  \ba{c}\la\obettildg_{\Del k}\obettil_{\Del k}\ra(t)-n^0_{\Del k}\\ 
        \la\oat_s\oatdg_s\ra(t)+n^0_{\Del k}\\
	\la\oat_s\obettil_{\Del k}\ra(t)-\mathrm{c.c.}
  \ea\rk)
  &=& \exp\lk[\lk(
    \ba{ccc} -\gam_{\Del k} & 0 & i\Ometil \\ 
             0 & 0 & i\Ometil \\
             -2i\Ometil & -2i\Ometil & -\gam_{\Del k}/2
    \ea\rk)(t-t_0)\rk]
    \lk(
    \ba{c}\la\obettildg_{\Del k}\obettil_{\Del k}\ra(t_0)-n^0_{\Del k}\\ 
          \la\oat_s\oatdg_s\ra(t_0)+n^0_{\Del k}\\
	  \la\oat_s\obettil_{\Del k}\ra(t_0)-\mathrm{c.c.}
    \ea\rk),
  \\
\label{eq83}
    \la\obettildg_{-\Del k}\obettil_{-\Del k}(t)\ra-n^0_{\Del k}
    &=&  \exp[-\gam_{\Del k}(t-t_0)]\lk(
         \la\obettildg_{-\Del k}\obettil_{-\Del k}\ra(t_0)-n^0_{\Del k}\rk).
\eea
Note that the operator ordering in the expectation values of the density operators is crucial.
This ordering may of course be changed in the final result using the commutation relations.
Thus (\ref{eq82}) shows that applying the basic equations of motion (\ref{eq74}) for the photon operators, at temperature $T=0$, where the equilibrium quasiparticle population $n^0_q$ vanishes, is only possible for their anti-normal-ordered products.
More generally, products of the operators $\oat_s$, $\obettil_{\Del k}$, and their respective hermitian conjugates have to be normal ordered in the sense, that $\oat_s$ has to be interpreted as a creation operator and $\oatdg_s$ as an annihilator, whereas the quasiparticle operators are taken in the usual sense.
This is due to the fact that our basic Hamiltonian in the approximation which leads to the equations (\ref{eq74}) is not excitation number conserving due to its pair production term $\propto\oatdg_s\obettildg_{\Del k}$.
For this reason we have already written in eqns.~(\ref{eq75}), (\ref{eq76}), and (\ref{eq79}) the crucial operator products in the required ordering.

The next step would be to derive corresponding equations for products of more operators, which can be done straightforwardly along the same lines.
However, as we are for the first interested in the zero-temperature case, we can restrict the effort by following the operator ordering rule stated above and using the basic time evolution equations (\ref{eq74}), (\ref{eq74b}).
In appendix \ref{appA} we show that this rule, applied to products of an even number of operators, in the Markov approximation leads to the correct time evolution of the expectation values at $T=0$.
\begin{figure}[tb]
\begin{center}
\epsfig{file={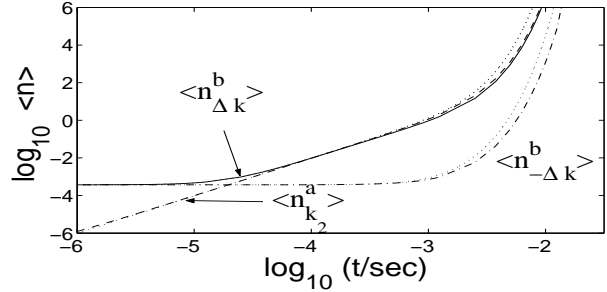},height=4cm,width=8cm,angle=0}
\caption{\label{fig2}Time dependence of the mean particle occupation numbers of the scattered photon mode, $\la\on^a_{k_2}\ra$ (dashed line), and of the modes of the recoiling atoms, $\la\on^b_{\Del k}\ra$ (solid line), $\la\on^b_{-\Del k}\ra$ (dashed-dotted line). The corresponding curves without decay are shown as dotted lines.}.
\end{center}
\end{figure}

\begin{figure}[tb]
\begin{center}
\epsfig{file={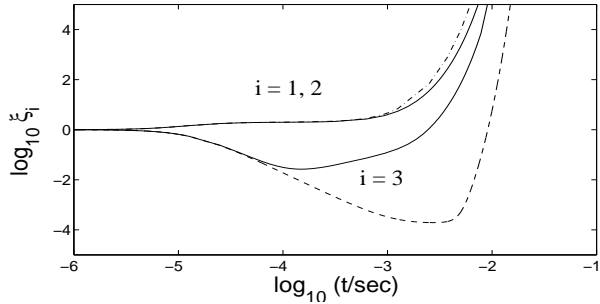},height=4cm,width=8cm,angle=0}\\[3mm]
\caption{\label{fig3} Time dependence of the squeezing parameters $\xi_i$, $i=1,2,3$, with (solid lines) and without quasiparticle decay (dashed and dash-dotted).}
\end{center}
\end{figure}

In the rest of this paper we present the results for the time evolution of the occupation numbers and squeezing in the same system as considered in \cite{GRB01}, but including quasiparticle damping.
We consider specifically a homogeneous condensate of $N=10^6$ $^{23}$Na atoms within a volume $V=10^{-8}\,$cm$^3$, which we choose such that the condensate density $n_0=10^{14}\,$cm$^{-3}$ is of the order which is typically observed in experiments. With $a=2.8\,$nm we thus have $k_0=2.65\cdot10^6\,$m$^{-1}$, $\hbar k_0^2/2m=(2\pi)\,1.5\,$kHz. 
The time dependence of the occupation numbers and the squeezing parameter is shown in Figs.~\ref{fig2} and \ref{fig3}. 
These results have been discussed in detail in \cite{GRB01}.
We plot the logarithm of the occupation numbers $\la\on^a_{k_2}\ra$, $\la\on^b_{\Del k}\ra$, $\la\on^b_{-\Del k}\ra$, for $\overline{\Del k}=\Del k/k_0=5$, i.e.~for coupling into the particle regime of the Bogoliubov spectrum, where $\ome_k\simeq \hbar k^2/2m$.
We have chosen a two photon Rabi-frequency of $\Ometil=1\,$s$^{-1}$. 
Initially the atomic occupation is $\la\on^b_{\Del k}(t=0)\ra=v_{\Del k}^2$ due to collisional interactions. 
Only for $t\gsim0.1\,$ms the photon and atom number occupations  are of the same magnitude, $\la\on^a_{k_2}\ra\simeq\la\on^b_{\Del k}\ra$.
For $\la\on^b_{\pm\Del k}\ra$ approaching $N_0=10^6$, the calculation yields non-physical results since we have neglected the depletion of $N_0$. 
The range of times where the calculation is valid may however be increased by choosing a smaller $\Ometil$.

The effects of the quasiparticle damping due to the finite width $\gam_{\Del k}$ become visible for $t\gsim\,1$ms.
The curves shown correspond to a width of $\gam_{5k_0}=2.8\cdot10^{-3}\ome_{5k_0}=7.0\cdot10^2\,$s$^{-1}$.
The damping affects all occupation numbers shown, also that of the photon mode, since the photon production rate depends on the occupation of the recoiling atomic mode.
A hypothetical stronger decay rate would result in a time evolution, where $\la\on^b_{\pm\Del k}\ra$ approach the initial value $v_{\Del k}^2$ for large times and $\la\on^a_{k_2}\ra$ grows at a considerably lower rate.

Fig.~\ref{fig3} shows the squeezing parameter $\xi_3$. 
For times $t\lsim10\,\mu$s it is close to $1$, i.e.~there is essentially no squeezing compared to the classical case of independent modes. 
This is because the relative atom-photon number variance is still dominated by the initial ``thermal'' variance of the atom mode population, $[\Del n^b_{\Del k}(t=0)]^2=\la\on^b_{\Del k}\ra(\la\on^b_{\Del k}\ra+1)=v_{\Del k}^2(v_{\Del k}^2+1)$, which, in the particle regime where $v_{\Del k}^2\ll1$, resembles the variance of a coherent state. (For recoil momenta in the quasiparticle regime, $\Del k<k_0$ the initial value of $\xi_3$ would be considerably larger than 1.)

For $t\gsim10\,\mu$s the squeezing starts to improve due to the atom-photon pair production, but only until $t\simeq0.1\,$ms.
It becomes clear that the quasiparticle damping causes the squeezing to deteriorate already at a considerably earlier time and larger value of $\xi_3$ compared to the loss-less case.
Decoherence due to quasiparticle damping is therefore a further important limiting factor.

In the context of spin squeezing, we found analytically that within the restricted Fock space of the modes $\pm\Del k$, $k_2$, we have $\la\oJ_1\ra=\la\oJ_2\ra\equiv0$ for all times. Thus the mean abstract spin vector $\la\oJv\ra$ points in 3-direction. Fig.~\ref{fig3} shows, that although the squeezing of all three spin components becomes worse for large times compared to the case of coherent modes, there is a relative squeezing of up to 2 orders of magnitude of the 3-component compared to the 1- and 2-components which are of equal size

\section{Conclusions}
\label{sec5}
In this paper we have studied the dynamics of interacting condensed Bose-gases at finite temperature.
We have set up a system of equations of motion for the operators and operator products which describe the fluctuations around the mean field and used the linear response approximation to describe the time evolution of the system.
We have focused on the dissipation in the time evolution of systems which initially are out of equilibrium, specifically for initial states with a single excited (quasiparticle) eigenmode.
As is well known, the decay of this mode is caused predominantly by interaction processes of the quasiparticles which involve a condensate atom and which can be classified as Landau or Beliaev processes.
We have reviewed the derivation of the general expressions of the quasiparticle decay rates for a trapped condensate at finite temperature and extended this to the case where the atoms in the initially excited mode are in a different internal state than the condensate atoms.
The interactions taken into account in this two-level case are collisions between atoms in different internal states which conserve the internal quantum number.
We calculated the coupling functions which enter the expressions of the Beliaev and Landau decay rates and found the same infrared $q^5$-proportionality of the Beliaev rate as for the single internal level case.

In order to gain a consistent description of the dissipative dynamics of densities and higher correlation functions in the long time limit we used the Heisenberg-Langevin approach.
In this way the time evolution equations of correlation functions including their correct equilibrium limit may be determined straightforwardly.
In this paper we are interested in the zero temperature dynamics of expectation values, which yield the number squeezing between different Fock modes of a system.
We were able to introduce and justify a regression theorem which we used to compute higher correlation functions using the basic equations of motion for the Fock operators only.

Specifically, we considered Bragg-scattering of photons off a finite homogeneous Bose-Einstein condensate of sodium atoms and studied the relative occupation number squeezing between the scattered photon and the recoiling atomic mode.
The parameter which quantifies the squeezing measures the variance of the number difference of the modes compared to its classical limit which is given by the sum of the mean occupations.
Hence our calculations required to determine the evolution of densities as well as mean values of products of four operators.
Our numerical results show that there is a substantial limitation of squeezing due to the collisional interactions between the atoms.
Compared to the approximation where quasiparticle interactions have been neglected the achievable squeezing is, in addition, considerably reduced by decoherence due to dissipation.

\section*{Acknowledgments}
I am especially grateful to Prof.~Keith Burnett who has helped much to make my stay in Oxford possible and joyful.
I would like to thank him, as well as Dr.~Thorsten K\"ohler, Dr.~Samuel Morgan, and David C.~Roberts for many lively discussions and valuable suggestions.
This work has been supported by the Alexander von Humboldt-Foundation, and the European Community under contract no.~HPMF-CT-1999-0023.

\begin{appendix}
\section{Operator ordering rule}
\label{appA}
In this appendix we derive the time evolution equations (\ref{eq82}), (\ref{eq83}), for the expectation values appearing on the right hand sides of (\ref{eq75},\ref{eq76}).
Moreover we justify our procedure, that, for $T=0$ and in the Markov approximation we may compute expectation values of normal ordered products of more than two operators by using the basic equations (\ref{eq74},\ref{eq74b}), without taking into account the Langevin operators.
Normal ordering here means, that $\oat_s$ is treated as a creation operator, and $\oatdg_s$ as an annihilator, whereas the quasiparticle operators are treated in the usual way.

In order to obtain the dynamical equations for the expectation values of the required products of two operators, we need to solve the following equations which are derived from (\ref{eq74},\ref{eq74b}).
\bea
\label{eq80}
  \frac{d}{dt}
  \lk(
  \ba{c}\la\obettildg_{\Del k}\obettil_{\Del k}\ra\\ 
        \la\oat_s\oatdg_s\ra\\
	\la\oat_s\obettil_{\Del k}\ra-\mathrm{c.c.}
  \ea\rk)
  &=& \lk(
    \ba{ccc} -\gam_{\Del k} & 0 & i\Ometil \\ 
             0 & 0 & i\Ometil \\
             -2i\Ometil & -2i\Ometil & -\gam_{\Del k}/2
    \ea\rk)
    \lk(
    \ba{c}\la\obettildg_{\Del k}\obettil_{\Del k}\ra\\ 
        \la\oat_s\oatdg_s\ra\\
	\la\oat_s\obettil_{\Del k}\ra-\mathrm{c.c.}
    \ea\rk)
    -\lk(
    \ba{c}\la\oFtdg_{\Del k}\obettil_{\Del k}\ra+\mathrm{c.c.} \\ 
          0 \\
          \la\oFtdg_{\Del k}\oatdg_s\ra-\mathrm{c.c.}
    \ea\rk),
  \\
\label{eq81}
    \frac{d}{dt}
    \la\obettildg_{-\Del k}\obettil_{-\Del k}\ra
    &=&  -\gam_{\Del k}\la\obettildg_{-\Del k}\obettil_{-\Del k}\ra
         -\lk(
	 \la\oFtdg_{\Del k}\obettil_{-\Del k}\ra+\mathrm{c.c.}\rk).   
\eea
The expectation values involving the Langevin operator determine the equilibrium values of the observables.
Using the formal integration of (\ref{eq74}), together with eqn.~(\ref{eq39}) for the two time correlation function, as well as eqns.~(\ref{eq13}), (\ref{eq14}), (\ref{eq36}) (which show that the expectation values of the Langevin operator with the Fock operators at time $t_0$ vanish), we find: $\la\oFtdg_{\pm\Del k}\obettil_{\pm\Del k}\ra=\gam_{\Del k}n^0_{\Del k}$, $\la\oFtdg_{\Del k}\oatdg_s\ra=0$.
This leads, upon integration of (\ref{eq80}), (\ref{eq81}), to the final result for the time evolution, eqns.~(\ref{eq82}), (\ref{eq83}).  

The crucial condition for the application of the regression theorem is the operator ordering in the expectation values of the density operators.
The general rule is that products of the operators $\oat_s$, $\obettil_{\Del k}$, and their respective hermitian conjugates have to be normal ordered.
This is meant in the sense, that $\oat_s$ has to be interpreted as a creation operator and $\oatdg_s$ as an annihilator, whereas the quasiparticle operators are taken in the usual sense.
In the following we sketch the proof that using the basic time evolution equations (\ref{eq74},\ref{eq74b}) and Wick's theorem for products of an even number of operators, in the Markov approximation leads to the correct time evolution of the expectation values at $T=0$.

The time evolution equation of a (in the above stated sense) normal ordered product of $n$ Fock operators, using the chain rule and eqns.~(\ref{eq74},\ref{eq74b}), involves terms containing the Langevin operator.
It is the expectation values of these terms which we have to consider and which need to vanish for the desired procedure to be justified.
Upon insertion of the formal solutions of eqns.~(\ref{eq74},\ref{eq74b}) for the Fock operators in these terms, their value is determined by expectation values of mixed products of Langevin operators at different times and Fock operators at time $t_0$.
The normal ordering of the initially considered operators carries through to a normal ordering of these mixed products.
Moreover, the terms involve time integrals over those Langevin operators which entered through the integrated eqns.~(\ref{eq74}), (\ref{eq74b}).
For the case of four operators which we need to compute squeezing, these integrals are e.g. of the form
\bea
\label{eqA1}
   &&\int_{t_0}^t\,dt'dt''dt'''\la\oFtdg(t)\oFtdg(t')\oFt(t'')\oFt(t''')\ra,
   \nonumber\\
   &&\int_{t_0}^t\,dt'\la\oFtdg(t)\oaldg(t_0)\oal(t_0)\oFt(t')\ra,
\eea
where $\oal$ is either of the operators $\oatdg_s$, $\obettil_{\pm\Del k}$.
When inserting (\ref{eq36}) we see that, at $T=0$, only those terms can be different from zero, which contain an operator $\og_{ij}(t_0)$ at the very left and one $\ogdg_{kl}(t_0)$ at the very right.
(If there is an $\oat_s(t_0)$ at the very left and/or an $\oatdg_s(t_0)$ at the very right, the expectation value may be broken down using Wick's theorem, and the problem reduced to the case of $n-2$ or less operators.)

The final step is to show that the remaining non-vanishing terms are zero in the Markov approximation.
This is done as before by assuming that the coupling functions (\ref{eq32}--\ref{eq34}) which multiply the integrals (\ref{eqA1}) vary only slowly over the range of modes, such that the sums over the mode indices result in functions which are strongly peaked at a single time and may be written as proportional to delta-distributions in time.
Integration over the time argument of one of the Langevin operators at the very right or left then immeditately yields, that the remaining terms are all proportional to $\del(\ome_{\Del k}+\ome_i+\ome_j)$.
Since we consider $|\Del k|\not=0$ only, the argument of the distribution is positive.
The expectation values in the equations of motion, which involve Langevin operators, therefore vanish in the Markov approximation at $T=0$, for the same reason as this is the case, with $n^0_q=0$, for (\ref{eq49}).\\

\end{appendix}

%


\begin{references}
\bibitem{fnpsiops}
Note that $\opsit$ is an operator whose hat we have suppressed for notatonal simplicity.
\bibitem{VectorNotation}
In the following, vectors $\kv$ are denoted by $k\equiv(k_x,k_y,k_z)$.
\end{references}
\end{document}